\makeatletter
\declare@file@substitution{revtex4-1.cls}{revtex4-2.cls}
\makeatother
\documentclass[twocolumn]{aastex631}
\hypersetup{linkcolor=red,citecolor=blue,filecolor=cyan,urlcolor=red}

\shorttitle{Magnetic waveguides and resonant cavities}
\shortauthors{Robert Sych}
\graphicspath{{./}{}}

\begin{document}

\title{First Imaging of Magnetic Waveguides and Resonant Cavities in Sunspots}

\correspondingauthor{Robert Sych}
\email{sych@iszf.irk.ru}

\author[0000-0003-4693-0660]{Robert Sych}
\affiliation{Institute of Solar-Terrestrial Physics SB RAS, Irkursk 664033, Russia}


\begin{abstract}

For the first time, we have determined the spatial distribution of magnetic waveguides and resonant cavities at different heights in the sunspot atmosphere. We applied a decomposition of time cubes of EUV/UV sunspot images obtained in the SDO/AIA temperature channels into narrowband components in the form of wave sources. The methods of pixelized wavelet filtering and oscillation mode decomposition were used. For all studied sunspots the presence  of selected bands in the spectra was shown. Each band corresponds to oscillations forming spatial waveguides in the volume of the sunspot atmosphere. The formation of waveguide bundles in the height from photospheric to coronal levels is shown. The regions of the waveguides with maximum oscillation power, where resonant cavities are formed, are identified. Their detection is an experimental proof of the theory of resonant layers, previously proposed to explain the presence  of significant harmonics in the oscillation spectrum. The different shapes of the cavities reflect the structure of the magnetic tubes along which the waves propagate. The distribution of sources in the height layers indicates the influence of the wave cutoff frequency caused by the inclinations of the magnetic field lines. We discuss the possibility of upward wave transport due to periodic amplification of the oscillation power in the detected cavities.

\end{abstract}
\keywords{magnetohydrodynamics (MHD) – methods: observational – Sun: atmosphere – Sun: oscillations – sunspots}

\section{Introduction}

		The wave dynamics associated with the 3-minute periodicity in sunspot atmospheres is a key phenomenon observed in the solar atmosphere at different wavelengths and atmospheric altitudes \citep{1966ApJ...143..917O, 1969SoPh....7..351B, 1972ApJ...178L..85Z, 1972SoPh...27...71G}. Although these waves are weak in the photosphere, they are amplified in the chromosphere and transition region. Observations show significant spatial fragmentation of the magnetic field within sunspots, coinciding with the sources of the oscillations, localized in the umbra and penumbra. These regions are associated with the footpoints of magnetic loops, where slow magnetoacoustic waves propagate from sub-photospheric levels into the corona. Several studies have investigated these phenomena, emphasizing the complex dynamics within sunspots and their impact on solar activity \citep{1974ApJ...189..359L, 2015LRSP...12....6K, 2016GMS...216..467S, 2023LRSP...20....1J}.
			
  The Fourier spectrum of the chromospheric umbra oscillations consists of multiple periodicities of about three minutes, with closely spaced peaks in the range $\sim4-20$~mHz. The dominant peak frequency of $\sim5.6$~mHz is associated with global oscillations, but the exact nature of the spectrum remains unknown. The frequency is thought to be influenced by the presence of resonators in the sunspot atmosphere and/or the acoustic cutoff period of the oscillation spectrum.
		
	There are two mechanisms to explain how oscillations in the lower solar atmosphere are amplified from photospheric to chromospheric heights. The first mechanism \citep{1981PAZh....7...44Z, 1990SoPh..128..353W, 2003ApJ...591..416C,2005SoPh..229..255Y, 2007AstL...33..622Z,2011ApJ...728...84B, 2018RAA....18..105Z, 2019AstL...45..177Z} is related to resonant cavities created by sharp temperature gradients. The second mechanism \citep{1977A&A....55..239B, 1984A&A...133..333Z} is related to the wave cutoff frequency, which can be varied by the divergence of magnetic field lines. This mechanism supports wave propagation in the solar atmosphere, allowing waves to reach coronal heights. A study by \cite{2010SoPh..266..349S} showed that photospheric oscillations can reach such heights if they are guided by tilted loop structures anchored in the sunspot umbra.
 
		The presence  of two types of resonators in the umbra - photospheric and chromospheric \citep{1981PAZh....7...44Z, 2008SoPh..251..501Z,  2011ApJ...728...84B}, with stratification of frequency ranges and spatial localization of oscillation sources is assumed. These can lead to the appearance of high-frequency oscillations about 10 mHz \citep{2012ApJ...746..119R, 2013A&A...554A.146K, 2016A&A...594A.101Y, 2018ApJ...856L..16W},  frequency drifts during wave propagation \citep{2012A&A...539A..23S}, and the formation of spiral wavefront structures in the umbra \citep{2014A&A...569A..72S, 2016ApJ...817..117S, 2019A&A...621A..43F, 2019ApJ...877L...9K, 2021RSPTA.37900180S}.
		    
		The presence of resonators in the sunspot atmosphere is indicated by the structure of nonlinearly propagating wave fronts in the chromospheric layers. Studies have shown \citep{2012A&A...539A..23S} that velocity fluctuations coincide with the appearance of high-frequency harmonics with periods of 1.2-1.5 minutes. During the development of 3-minute wave trains in sunspots, frequency drifts with power variations are observed at all height levels of the atmosphere. Drift rates range from 5 mHz/hour in the photosphere to 13 mHz/hour in the corona. These frequency drifts are thought to be caused by the passage of waves through isolated resonant layers that affect their propagation. Another resulting from the occurrence of strong oscillations in sunspot inhomogeneities is the appearance of separate details of wave fronts with different cutoff frequencies \citep{2014A&A...569A..72S}, forming their spiral shape in the 3-minute spectral band.
		    	
			  In 2020, \citet{2021NatAs...5....5J} made spectropolarimetric observations of a large sunspot and identified isolated spectral bands where the energy of Doppler velocity fluctuations was localized. They divided the spectral windows into three bands: low-frequency oscillations below 4 mHz, broad spectral peaks in the $\sim5-17$~mHz range, and high-frequency oscillations in the $\sim18-27$~mHz range. Statistical analysis showed a correlation between the steepness of the slope of each oscillation energy band and the distance to the sunspot center. A high frequency component was detected at a frequency of about 20 mHz. All this points to the possible presence of resonant cavities at different distances from the umbra center, where the amplitude of the oscillations increases and significant harmonics are formed.
		
			In our work we have investigated wave processes throughout the sunspot atmosphere using UV/EUV observations of different layers of the solar atmosphere. We continue our recent study \citep{2024MNRAS.529..967S}, where for the first time two-dimensional stable regions associated with waveguides in the sunspot umbra were observed in separate SDO/AIA temperature channels. The angular size of these regions, identified as resonant cavities, was shown to vary with frequency. These regions coincided with the harmonics of the main oscillation mode and frequency drift. Their spectral slopes varying as a function of the location of the source in the umbra. It was found that at 304~{\AA} the high frequency sources of oscillations were localized at the footpoint of the magnetic bundle. The distribution of wave sources in the umbra allowed us to reveal the presence of resonant cavities at separate wavelengths, amplifying the oscillations at the selected harmonics.
			
In our work, we assume that the period of the propagating waves mainly depends on the cutoff frequency, which varies as a cosine function of the inclination of the magnetic waveguides to the solar normal \citep{2012ApJ...746..119R, 2014A&A...561A..19Y}.  The paths of the propagating waves in the sunspot volume can be obtained by determining the shape of these sources at different frequencies and their localization in space and height at different SDO/AIA temperature channels.

In the first time we imaged the three-dimensional spatial structure of magnetic waveguides over the entire height of the sunspot atmosphere and analyzed their spatial structure at different heights. We used spectral processing techniques for a number of sunspots such as pixelized wavelet filtering (PWF) \citep{2008SoPh..248..395S} and pixelized mode decomposition (PMD) \citep{2021RSPTA.37900180S} to obtain spectra and images of narrowband sources. The study confirmed the presence  of resonant cavities in certain ranges of oscillation periods at all sunspot heights, whose angular size varies with height above the solar surface. It is shown that the cavities are part of the waveguides where the amplification of the oscillation power and the formation of selected harmonics of the spectrum occur. Depending on the frequency range of the considered waves and the magnetic field strength, the mutual position of the boundaries of the resonant cavities and their changes can influence the resonant frequencies.

The paper is organized as follows: Section 1 presents the topic of the paper; Section 2 describes the observational data and methods of processing; Section 3 presents the data analysis and the obtained results; Section 4 considers the physical processes associated with the formation of resonant cavities; Section 5 draws conclusions on the obtained results.

\section{Observations  and Methods} 
    
		To study the structure of the oscillation sources, we analyzed one-hour EUV/UV emission observations of six single sunspots in the active groups NOAA 11131 and NOAA 11133 (December 09, 2010, 03:00 UT), NOAA 13549 (January 17, 2024, 04:00 UT), NOAA 13586 (February 20, 2024, 04:00), NOAA 13314 (May 23, 2023, 04:00 UT), and NOAA 13719 (June 23, 2024, 04:00 UT). The sunspots have simple symmetric shape with areas 330, 120, 180, 160, 20, 150 in millionths of the solar disk area respectively. The big group NOAA 11131 was chosen both because of the active study of oscillations \citep{2012ApJ...756...35R, 2014A&A...569A..72S, 2014AstL...40..576Z} and the resonant cavities found at selected wavelengths \citep{2024MNRAS.529..967S}. The groups were close to the central meridian without being affected by the effect of projection onto the solar disk. We studied the observational data with the AIA (Atmospheric Imaging Assembly) \citep[AIA;][]{2012SoPh..275...17L} on board the SDO (Solar Dynamics Observatory) \citep[SDO;][]{2012SoPh..275....3P} in the wavelength ranges 1700~{\AA}, 1600~{\AA}, 304~{\AA}, 171~{\AA}, 193~{\AA}, and 211~{\AA}. The choice of temperature channels is related to the study of sunspot oscillations over a wide range of heights where their observed, from the temperature minimum (1700~{\AA}) to the hot corona (211~{\AA}).
			
    The AIA images the entire solar disk in several EUV/UV wavelength bands almost simultaneously. The pixel size of the images is 0.6 arcsec, and the temporal resolution for all wavelengths was chosen to be 24 seconds in order to compare intensity variations over time. The duration of the observations and the temporal resolution allowed us to study variations in the period range from 48 second to 20 minute or 0.8-21 mHz in frequency.  The size of the study area was $41\times41$ arcsec, completely covering the sunspot. This provided spatial information about the sources of oscillations in both the umbra and the penumbra.
			  
		For calibration, the acquired AIA images were preprocessed using standard "aia-prep.pro" procedures and have the same spatial resolution of the order of 1 arcsec. The Solar Soft library software - \href{https://sohowww.nascom.nasa.gov/solarsoft/}{https://sohowww.nascom.nasa.gov/solarsoft/} was used to remove the differential rotation of sunspots during the observation. The temporal variations of the emission brightness of each pixel were studied using two spectral methods: PWF filtering \citep{2008SoPh..248..395S} and PMD decomposition \citep{2021RSPTA.37900180S} of signals. The use of PMD analysis allowed us to study the amplitude distributions of the oscillations with frequency, to obtain their energy spectrum with high spectral resolution, and to identify separate enhancement bands of the resonance signal. Using PWF analysis, we obtained the three-dimensional power distribution of oscillations in space in the form of waveguides, studied the structure of resonant cavities in them and followed their changes in height.

\section{Analysis and Results} 

The analysis profiles of UV/EUV emission oscillations in space for one SDO/AIA temperature channel gives the property of only a thin slice of the sunspot atmosphere in height. Oscillations are present in different temperature channels. Since the measured atmospheric temperature depends on the height above the visible surface of the sunspots, we studied the oscillations over a wide range of wavelengths recorded by SDO/AIA at 1700~{\AA}, 1600~{\AA}, 304~{\AA}, 171~{\AA}, 193~{\AA}, and 211~{\AA} ~ by creating time cubes of the images. The temporal resolution between frames was 24 seconds for all channels, and the duration of the observations for the active groups was 60 minutes. The combined use of wave propagation data for all channels allowed us, for the first time, to obtain a three-dimensional distribution of the oscillation sources in both space and height.

		\begin{figure*}
		\centering
\includegraphics[width=18cm]{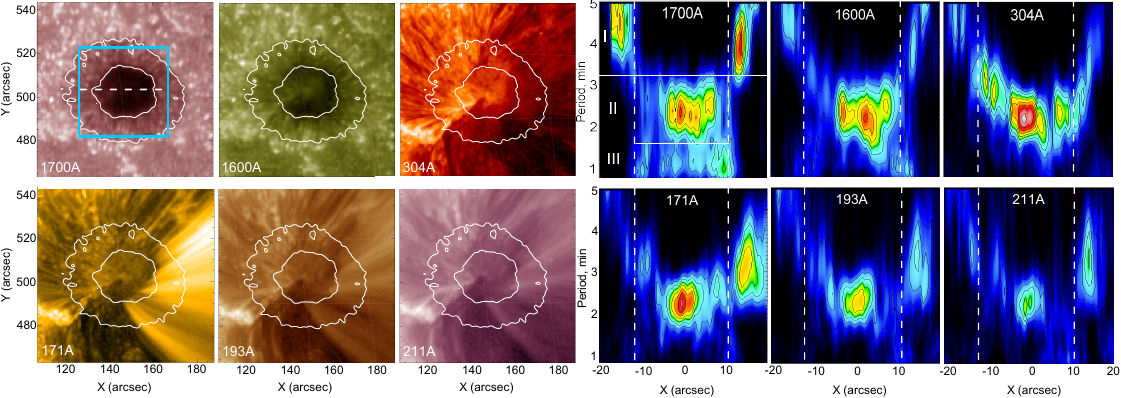}
\caption{\textit{Left panels}: Images of sunspot active group NOAA 11131 on 09 December 2010, 02:59:44 UT at 1700~{\AA}, 1600~{\AA}, 304~{\AA}, 171~{\AA}, 193~{\AA}, and 211~{\AA} channels. Thick white contours indicate the optical boundaries of umbra and penumbra. The region of interest (ROI) is indicated by a blue square. The horizontal broken line show the slice for 1D spectral analysis.  Spatial coordinates are given in heliocentric arcseconds, while image magnitudes are given on a logarithmic scale to improve visual clarity. \textit{Right panels}: Distribution of power oscillations as 1D period-coordinate plots for different wavelengths. The vertical lines show the umbra boundary. The horizontal lines and numbers at 1700~{\AA} show the selected period bands where the resonant cavities are found.
The periodicity is given in minutes.}
\label{fig:1}
\end{figure*}

\subsection{One-dimensional distribution of power oscillations}

Figure \ref{fig:1} (left panels) shows images of the studied sunspot active group NOAA 11131 obtained on December 09, 2010 in different SDO/AIA temperature channels.  We see a symmetric sunspot with maximum depression at the level of the temperature minimum at 1700~{\AA}. The umbra and penumbra boundaries are indicated by white contours. The brightness of the images is presented in a logarithmic scale. As the  height above the surface increases, a bright V-shaped magnetic structure connecting the main sunspot with the trailing part of the group begins to form, most clearly visible at wavelength 304~{\AA}.  A similar structure of the group was previously studied  in \cite{2012ApJ...756...35R, 2024MNRAS.529..967S} and is associated with the predominance of low loops in the eastern direction. The power of sunspot oscillations is maximal at the chromosphere level. In the corona, the spatial arrangement of magnetic loops in channel 171~{\AA} ~has an asymmetric distribution with respect to the central part. High loops, anchored in the center of the sunspot umbra and closing westward to the outer background regions, predominate. In the hot temperature channels 193~{\AA} ~and 211~{\AA} ~the brightness of the loops decreases. The blue square shows the region of interest. The magnetic field of the sunspot about 2 kGs.
		
		      To obtain a one-dimensional distribution of the period oscillations in the sunspot over an hour interval, we used the variations of all spatial points of the image cube over time on a horizontal slice passing through the umbra centre, shown by the broken horizontal line at 1700A~{\AA} panel~(Fig.~\ref{fig:1}).  The spectral analysis of the oscillations was performed using a PWF method \citep{2008SoPh..248..395S}. Using spectral filtering, it is possible to obtain 1D and 2D narrowband images of the oscillation sources in the period band, calculate their amplitude, power and phase characteristics, and track their changes over time. The use of SDO/AIA for multiple channels made it possible for the first time to obtain height distributions of oscillation parameters and to create volumetric images of waveguides. 
							
			The obtained one-dimensional plots of the dependence of the oscillation period on the distance to the umbra center are shown in Fig.~\ref{fig:1}, right panels. Similar plots were obtained for all SDO/AIA wavelengths. The vertical dashed lines indicate the umbra boundaries.
			
			\begin{figure*}
			\centering
\includegraphics[width=18cm]{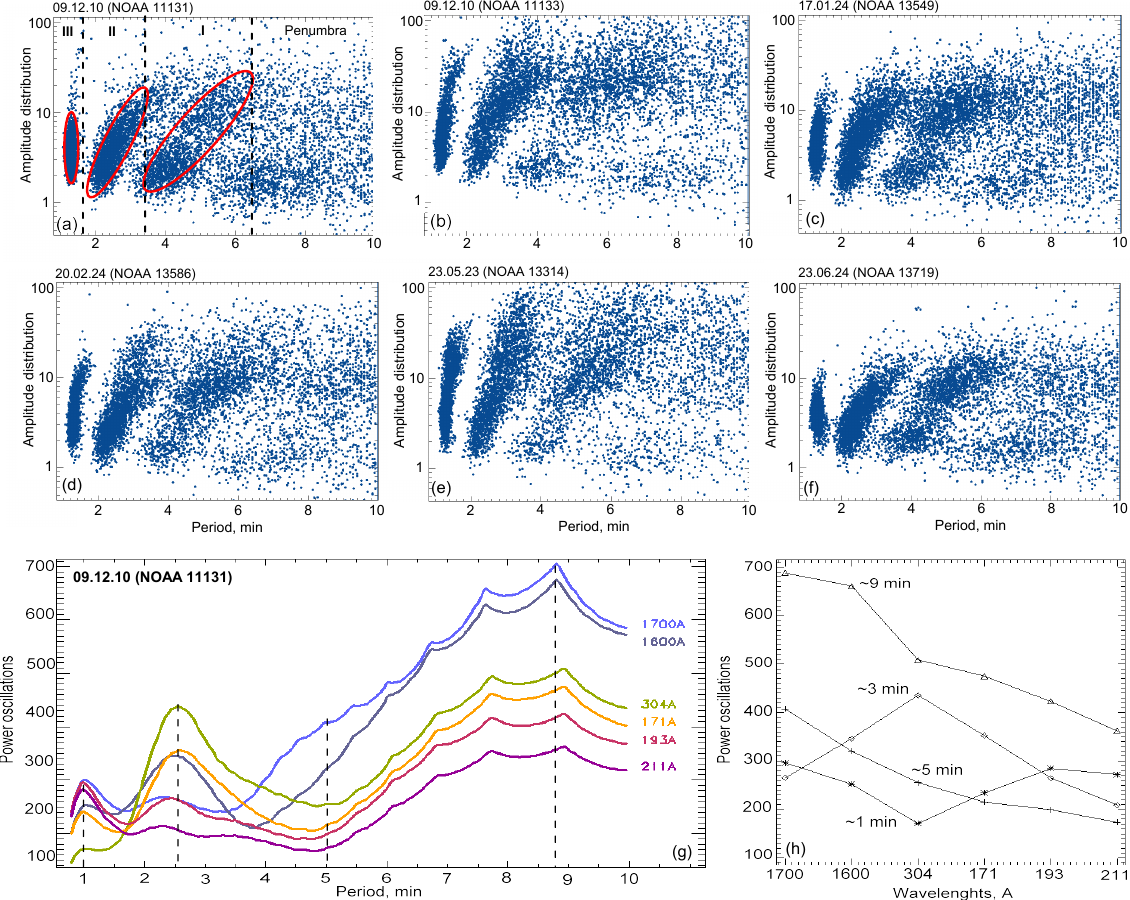}
\caption{Power spectra of sunspot oscillations obtained with different spectral techniques. \textit{Upper panels}: PMD spectra of six active groups - NOAA 11131 (a), NOAA 11133 (b), NOAA 13549 (c), NOAA 11586 (d), NOAA 13314 (e), and NOAA 13719 (f) at 1700~{\AA}. Vertical lines separate the spectral regions labeled I, II, III. Red ellipses show the amplitude distribution with certain periodicity. \textit{Bottom panels}: (g) The color profiles show the power oscillations of sunspot group NOAA 11131 at different temperature channels 1700~{\AA}, 1600~{\AA}, 304~{\AA}, 171~{\AA},193~{\AA}, 211~{\AA} obtained with use the PWF digital technique. The vertical dashed lines indicate the period localization of the maximum harmonics. (h) Dependence of power oscillations with UV/EUV wavelengths for selected periodicities.}
\label{fig:2}
\end{figure*}
			     	
		It can be seen that different parts of the sunspot show different signal periodicity depending on their distance from the sunspot center and the height of the emitting layer. At the bottom, at the level of the temperature minimum (1700~{\AA}), we observe an X-shaped structure of sources with three distinct period bands about 1-2, 2-3, and 3-5 minute, indicated by horizontal lines. All sources are separate and have a beginning and an end in terms of periods. The spectral bands are labeled I, II, and III, respectively. The high frequency band III shows the convergence of elongated spatial structures towards the center and the formation of the brightest part with a periodicity of 3-minute (band II). The mostly small band III sources are anchored at the inner umbra boundary. Sources of the low frequency part of band I are located in the penumbra with the formation of elongated divergent bright structures. Their brightness is maximum at this wavelength.
				
		At 1600~{\AA} there is also a convergence of oscillation sources towards the center, but their brightness decreases. In band II more elongated sources appear, gradually filling the umbra of the sunspot. The sources in the penumbra are fading. 
		
		The plot at 304~{\AA} ~shows a change in the distribution of the spectral power of the oscillations. There is a transition of the sources from X to V shape. Band II or so called 3-minute oscillations dominate. The band occupied by them widens with the formation of many tilted elongated structures occupying the whole umbra. There is a single footpoint of the forming bundle. From the right side in the penumbra (western part of the sunspot) a bright pattern is formed, which is a reflection of the eastward amplification of the oscillation power propagating along the open field lines. The power reaches a maximum at 171~{\AA} ~with a decrease at hotter altitudes at 193~{\AA} ~and 211~{\AA}. Note that the angular size of the 3- minute oscillation source (band II) gradually decreases as the emission layer extends into the corona. Similarly, its brightness decreases.
		
The obtained data confirm earlier results from one-dimensional spectroscopic observations \citep{2020NatAs...4..220J, 2021NatAs...5....2F} and two-dimensional imaging \citep{2024MNRAS.529..967S}, which revealed sunspot regions in the form of resonant cavities with amplification of narrowband oscillations. This suggests a complex internal structure of waveguides in the sunspot atmosphere. 
		
		\subsection{Oscillations spectra}
				
	 The use of PMD analysis allowed us for the first time to obtain the energy distribution in the spectra of sunspot oscillations in the form of extracted frequency bands. To obtain the spectra, we used the decomposition of the signals of each pixel during observation into inner functions. The PMD spectrum is the distribution of the oscillation amplitudes of each of the 6400 spatial image pixels over time as a function of period.
				
				Figure \ref{fig:2} (upper panels) shows the calculated PMD spectra of sunspot oscillations for all studied groups NOAA 11131 (a), NOAA 13314 (b), NOAA 11133 (c), NOAA 13586 (d), NOAA 13549 (e), and NOAA 13719 (f) at the 1700~{\AA}~ temperature minimum level. At this level, three bands of periods separated by vertical lines are most clearly identified in the umbras for all groups. 
				
				All of them correspond to the previously detected spatial brightening in the period-coordinate plots (Fig.~\ref{fig:1}). A broadening of the frequency band in amplitude near the maxima of the oscillations is observed, as well as their inclinations (drifts), are indicate by ellipses (Fig.~\ref{fig:2}a). The level of the period drift is maximal for the long-period oscillations in the penumbra and minimal for the high-frequency component in umbra. Each local pixel makes unique amplitude-frequency contribution to the total spectral bandwidth of the sunspot oscillations.  A continuous spectrum with a single maximum, such as a 3-minute period, is not observed.  The values of the spectral bands found are as follows: III ($\sim12-22$~mHz), II ($\sim5-12$~mHz), and I ($\sim1-5$~mHz).  The multiplicity of the  obtained frequency bands indicates on possible relation to harmonics of the global p-mode oscillations.
				
				\begin{figure*}
				\centering
\includegraphics[width=18cm]{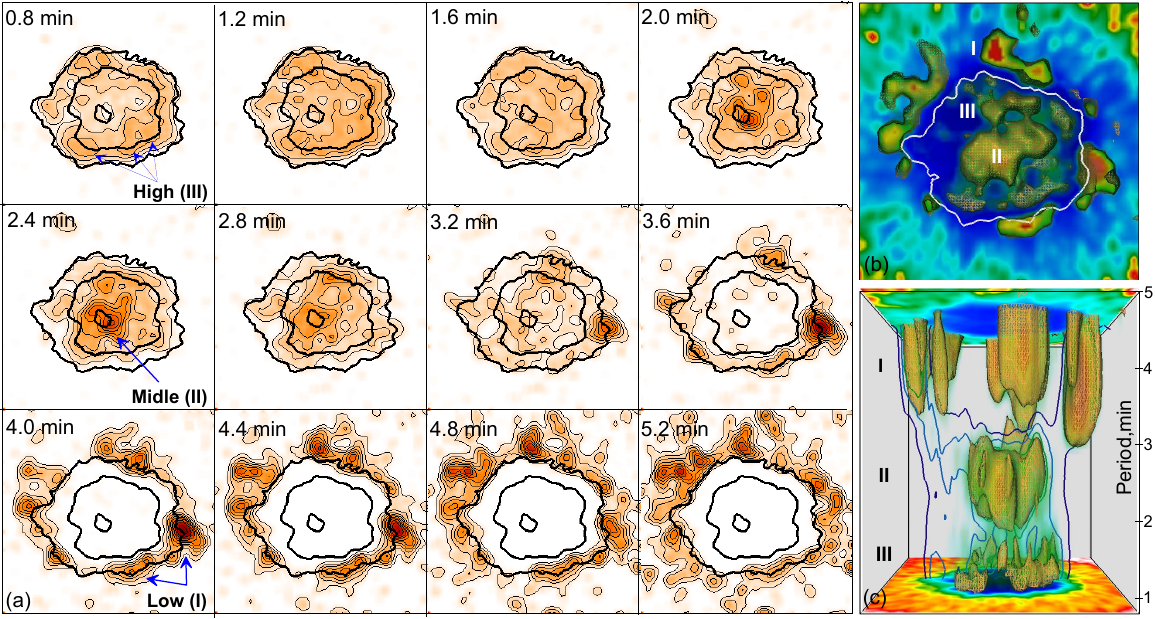}
\caption{(a) Narrowband images of sunspot oscillations in NOAA 11131 with 0.8-5.2 minute periodicity at 1700~{\AA} wavelength. Power oscillations are indicated by filled brightness levels from 0.1 to 0.9.  Maxima of the sources for different spectral regions are indicated by the blue arrows. Top (b) and side (c) views shows the 3D distributed narrowband oscillation sources (waveguides). The lower part of image (c) is a white-light sunspot image, the upper part and background on (b) is an image obtained at 1700~{\AA} wavelength. Central vertical cuts (c) are shown with blue contours. The spatial position of significant spectral regions is indicated by numbers. The period scale is in minutes. The thick contours in (a) indicate the borders of the white-light umbra at 0.01, 0.1, and 0.5 of maximum intensity.}
\label{fig:3}
\end{figure*}
			
			The lower panels show profiles of global PWF spectra obtained for oscillations at different heights (temperature channels) from 1700~{\AA} ~to 211~{\AA} ~(Fig.~\ref{fig:2}g) for the NOAA 11131 active group. They represent the integration of narrowband oscillation power sources over the entire sunspot for separate periods obtained by PWF method and ranging from 0.8 to 10 minute.  The higher power levels in the spectrum indicate a greater number of detected narrowband sources.  Each profile is shown with a different color depending on the wavelength. The detected peaks on the PWF spectra are indicated by vertical dashed lines. The peaks appear differently at different wavelengths and are the average analog of the power of individual harmonics on the PMD spectra (Fig.~\ref{fig:2}a). The minimum in the oscillation spectrum is the high-frequency harmonic near the 1-minute periodicity, while the maximum is the low-frequency component near the 9-minute periodicity.
					
		We have plotted profiles of the wavelength dependence of the oscillation power at the peak periods (Fig.~\ref{fig:2}h). Different dependencies are observed for different periods. It can be seen that the power of all oscillations decreases as one moves upward in the sunspot atmosphere. The high frequency part near the 1-minute period is present at all wavelengths up to the 211~{\AA} ~hot corona. Its power is maximal downwards, at the level of the temperature minimum (1700~{\AA}), and minimal at the level of the chromosphere (304~{\AA}). In contrary, the three-minute component increases and reaches a maximum at the level of the chromosphere (304~{\AA}), followed by a decrease in the corona. The low frequency components with periods about 5 and 9 minute are maxima in the lower layers of the atmosphere and also show a decrease in oscillation power at coronal levels.

	\subsection{Volume structure of the sunspot waveguides}

The presence of sources in sunspots where the amplification of the oscillations occurs on the selected harmonics is indicated by the spectra constructed using different signal processing methods (Fig.~\ref{fig:2}).  The characteristics of these sources vary depending on their height above the solar surface.  In order to obtain information about these sources, we used the performing of narrowband oscillation sources for the selected frequency bands using pixel-by-pixel wavelet filtering of each pixel of the images.

Figure \ref{fig:3}a shows the narrowband oscillation sources at the level of the temperature minimum (1700~{\AA}) in the period range of 0.8-5.2 minutes. The thickened lines show the brightness levels of the umbra boundaries at 0.01, 0.1, and 0.5 of the quiet Sun level. All highlighted components show the maximum power of the signal variation. We can see that in the III spectral range with short periods of ~ 0.8-1.6 minutes, the entire sunspot umbra is filled with these variations. The small sources are mostly clustered near the inner umbra boundary and are indicated by the blue arrows. As the period of the oscillations increases, sources in the II spectral range are formed in the center of the umbra with growing angular size and then disappear with increasing period. The low frequency part of the spectrum (I region) is represented by large, elongated sources with maximum brightness localized in the penumbra.

The obtained data on the appearance and disappearance of 2D narrowband sources in space as a function of the oscillation period reflect their discreteness and are related to the behavior of the previously detected harmonics in the spectra (Fig.~\ref{fig:2}) and the 1D distribution of the oscillation power in space (Fig.~\ref{fig:1}, right panels).  We have presented the obtained result as a 3D cube of the oscillation power distribution in space-period coordinates (Fig.~\ref{fig:3}b, top view) and (Fig.~\ref{fig:3}c, side view). The numbers I, II, III indicate the detected spectral bands and the corresponding oscillation sources. The base of the cube is the image of a sunspot in the optical range. The upper part of the sunspot is shown at wavelength 1700~{\AA}. The brightness of the volume parts depends on the power of the oscillations. The white contour shows the umbra boundary. The vertical section of the umbra center, shown as blue contours in Fig.~\ref{fig:3}c, indicates the presence of low-power background oscillations in volume of the cube. The waveguides, occupying separate I, II, and III period bands, together form a thick bundle of magnetic tubes anchored in the umbra. In the penumbra it begins to fragment into individual waveguides and to diverge with appearance tilts to the solar normal.

The separation of the oscillation extrema (Fig.~\ref{fig:3}b, c) revealed the presence  of selected narrowband waveguides that trace the motion of slow magnetoacoustic waves in the volume of sunspot atmosphere. Here the oscillation power reaches its maximum values. These waves move along the magnetic tubes. Their localization by periods coincides well with the previously detected spectral bands. Down in the III high-frequency band, the sources are associated with numerous short tubes that form a circular bundle converging toward the center in a narrow period range. The peripheral distance between them is minimal. The diameter of these tubes is about 3-5 arcseconds. Their footpoints are anchored in the inner part of the umbra (Fig.~\ref{fig:3}b, top view). In band II we observe the enlargement of these tubes due to their merging and widening with the formation of a vertical thick bundle in the center of the umbra. The oscillation power maxima are located in the central part and are indicated by the number II in Fig.~\ref{fig:3}b. The beginning and the end of this pattern in terms of periods completely coincide with the appearance of features in the obtained spectra (Fig.~\ref{fig:2}). The occupied range of oscillations is extended compared to the high frequency part. At low frequencies (band I), divergent thick waveguides begin to form along the outer perimeter of the penumbra. There is a deviation of them from the vertical position with an increase in angular size up to 10-15 angular seconds.  They are all extensions of the bundle visible in the spectral band II.
   	
		The obtained distributions of the oscillation power in the spectra (Fig.~\ref{fig:2}) and the sunspot volume at the 1700~{\AA} ~wave (Fig.~\ref{fig:3}) indicate the presence of selected spectral regions where amplification of the oscillation power occurs. We assume that these regions are connected with resonance cavities which previously predicted and studied  in \cite{2021NatAs...5....5J, 2024MNRAS.529..967S}.  The obtained result shows the reality of these regions in the form of isolated sections of magnetic waveguides distributed both in space and in period
oscillations.
		
		\begin{figure*}
\centering
\includegraphics[width=18cm]{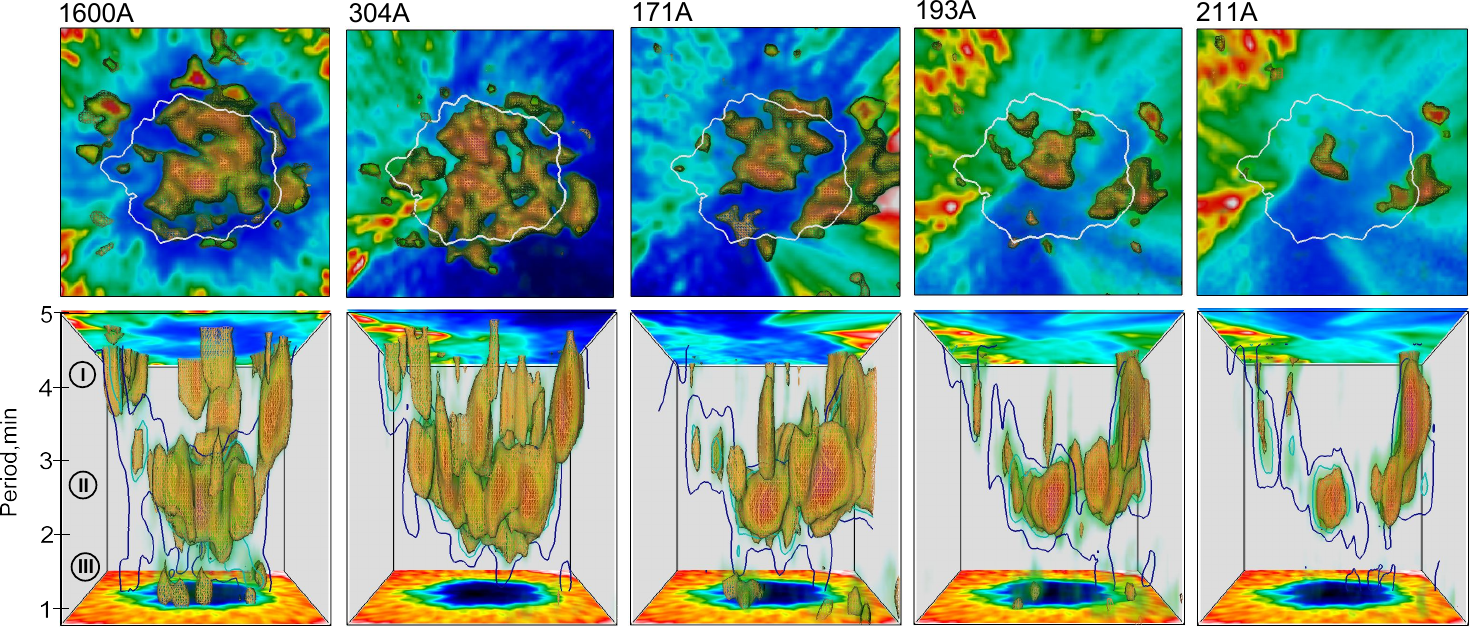}
\caption{\textit{Upper panels}:  Top view of the 3D narrowband distribution of sunspot oscillations in NOAA 11131 with 0.8-5.0 min periodicity at different SDO/AIA temperature channels 1600~{\AA}, 304~{\AA}, 171~{\AA}, 193~{\AA}, 211~{\AA}.  The background is the original UV/EUV intensity distribution.   \textit{Bottom panels}:  Similar images to the upper panels, but for the side view. Central vertical sections of the data cube are shown with blue contours. The spatial position of significant spectral regions is indicated by numbers at 1600~{\AA} panel. The period scale is in minutes. The white contour indicates the umbra boundary.}
\label{fig:4}
\end{figure*}

		Since the 1700~{\AA} wave observations cover only a narrow part of the sunspot atmosphere below, it is necessary to trace all SDO/AIA wavelengths to study how the detected resonant cavities are distributed in height down to the coronal levels. For this purpose we used the channels 1600~{\AA}, 304~{\AA}, 171~{\AA}, 193~{\AA}, and 211~{\AA}, where we followed the oscillations in the PWF spectra (Fig.~\ref{fig:2}g) of the active group NOAA 11131. The period distribution cubes of the power oscillations were prepared for all wavelengths, similar to the 1700~{\AA}. Figure \ref{fig:4} shows the result, where the top panels show the top view and the bottom panels show the side view of the obtained data cubes.
		
		\begin{figure*}
		\centering
\includegraphics[width=18cm]{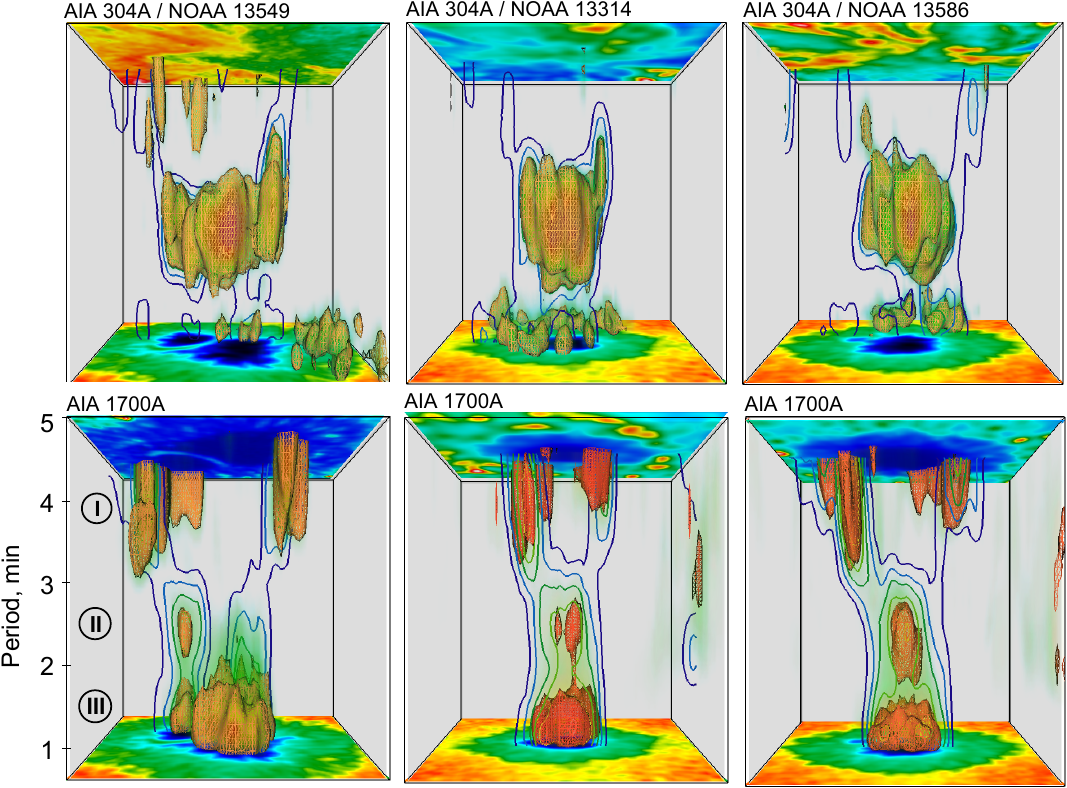}
\caption{3D structure of narrowband oscillation waveguides of sunspot active groups NOAA 13549, NOAA 13314, NOAA 13586 with 0.8-5.0 min periodicity in the chromosphere (304~{\AA}, upper panels) and photosphere (1700~{\AA}, bottom panels).}
\label{fig:5}
\end{figure*}

 With increasing altitude, at the level of the upper photosphere 1600~{\AA}, a similar pattern of period distribution of the oscillations as at 1700~{\AA} is observed, with increasing thickness of the tube bundles in band II and their further broadening and merging. They are basically a continuation of the tubes detected earlier at 1700~{\AA}. New ones begin to appear away from the central part of the umbra. The low frequency part is mostly localized in the same places. The number of tubes and their power decreases in the high frequency part.
  	
		In the chromosphere at 304~{\AA}, the number of new tubes with 3-minute oscillations increases sharply, filling almost the entire umbra. They begin to form separate thick bundles, which merge and begin to exit into the penumbra region, occupying the Ist low-frequency spectral band. There is a disappearance of the old bundles, previously visible at 1700~{\AA} and 1600~{\AA}, and their stratification into thin tubes. Some of these are located at the footpoints of the low loops anchored to the trailing part of the group.  Instead of multiple converging high frequency loops, we see formations of a single bundle root anchored in the center of the umbra.
				
			 At the level of the quiet corona, the 171~{\AA} component in band II is represented by the merging of tubes into several thick bundles in the center. There is a redistribution of oscillation power to the west. This indicates the absence of significant oscillations in the low loops in the eastern direction, anchored on the trailing part.  The low-frequency component is mainly localized at the footpoints of open coronal loops. Its magnitude varies with the number of closed loops. Each magnetic loop is connected to a separate waveguide along which the waves propagate. Traces of high-frequency oscillations are detected at the footpoints of the high loops. 
	
     At the 193~{\AA} ~and 211~{\AA} ~hot coronal levels, the number of oscillation sources decreases sharply. We see that the 3- minute oscillations in band II can penetrate high into the corona. Their sources are vertical tubes in the center of the umbra, whose angular size decreases as they reach the corona.  In the penumbra, elongated tubes associated with the footpoints of hot coronal loops dominate, along which running waves propagate.
						
					The obtained position of narrowband sources in the sunspot active group of NOAA 11131 showed the distinct regions of magnetic waveguides at different heights where the amplification of oscillations occurs. To verify our results and obtain statistics, we studied five other groups, NOAA 13549, NOAA 13314, NOAA 13586, NOAA 11133, and NOAA 13719 with unique magnetic configurations. Oscillation distributions were obtained by creating 3D space-period data cubes for different wavelengths, similar to the NOAA 11131 described above. 	Figure \ref{fig:5} shows sunspots in NOAA 13549, 13314, and 13586 active groups at the chromospheric (304~{\AA}, upper panels) and photospheric (1700~{\AA}, lower panels) levels.
			
			All 3D cubes show X- and V-shaped bundles with the formation of selected waveguide parts in the form of photospheric and chromospheric resonators. The angular umbra size and magnetic field strength of the considered sunspots are smaller than in the developed sunspot of the NOAA 11131 group. Vertical fields dominate, with no significant divergence of the field lines. This leads to the fact that at 1700~{\AA} the high frequency details in the form of a loops bundle are vertical and completely fill the umbra without significant divergence.	
			
					\begin{figure*}
					\centering
\includegraphics[width=18cm]{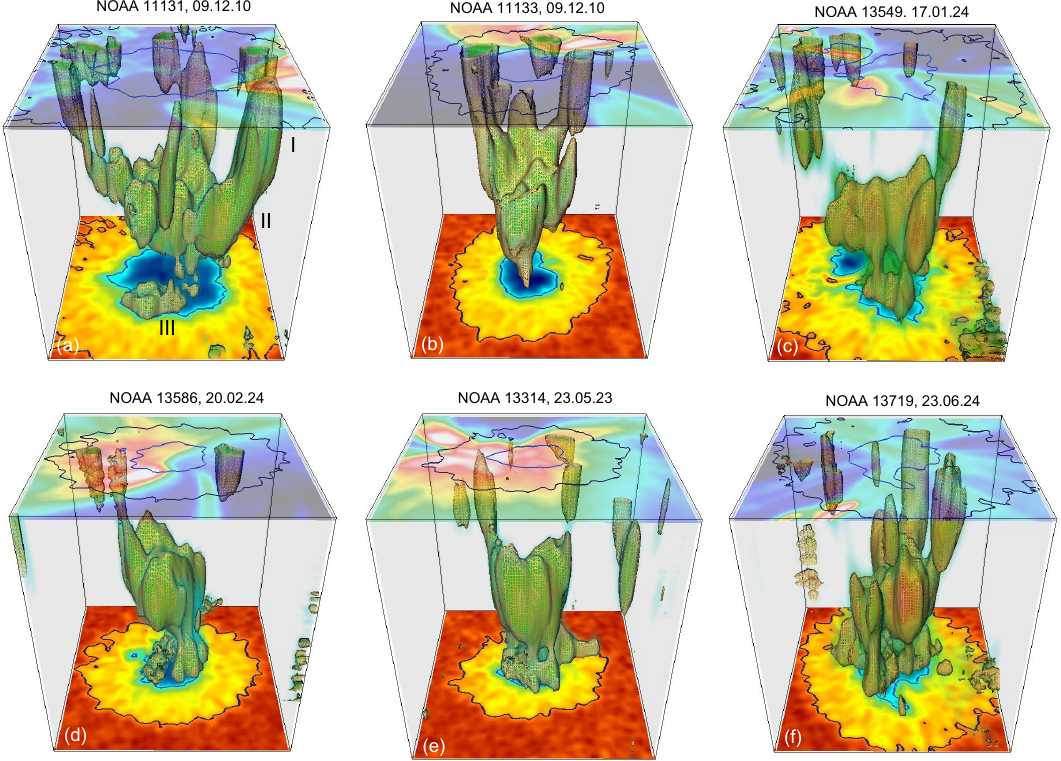}
\caption{The composite 3D height structure of waveguides with 0.8-5.0 min periodicity in sunspot active groups NOAA 11131 (a), NOAA 11133 (b), NOAA 13549 (c), NOAA 13586 (d), NOAA 13314 (e), and NOAA 13719 (f).  The lower part of the images is a white-light image of the sunspot, the upper part was obtained at 171~{\AA} wavelength. The black contours show the umbra and penumbra boundaries.}
\label{fig:6}
\end{figure*}

     We can observe the formation of two bundles in sunspot group NOAA 13549, anchored in different umbras separated by a light bridge.  One bundle is observed to amplify and form a resonant cavity in II-nd band, while it is not observed in the other bundle. This indicates a difference in the physical parameters of the waveguides where resonant wave amplification is possible. The formed thin bundles have a simple spatial structure in the IIth spectral band, without significant details. Each bundle ends with an exit into the penumbra with the formation of a low-frequency source.

In the chromosphere, all groups show an increase in the angular size of the 3-minute chromospheric resonator in the form of closely spaced tubes.  The high-frequency part shows a number of tubes, indicating their upward penetration with stratification of the dense bundle into separate components.  The number of low-frequency sources is minimal, indicating a vertical field with no significant field tilt.

The obtained separate spatial distributions of the oscillation power at selected SDO/AIA temperature channels or heights (Fig.~\ref{fig:4})~allows us to prepare the full spatial structure of the waveguides from the photosphere to the corona. To prepare the volume waveguides we used of the two-dimensional distributions of narrowband sources at temperature channels 1700~{\AA}, 1600~{\AA}, 304~{\AA}, 171~{\AA}, 193~{\AA}, and 211~{\AA}. The order of the channels used corresponds to their power oscillations level (Fig.~\ref{fig:2}) and height above the photosphere. For each period in the band 0.8-5 min we performed  the cyclic preparation of narrowband oscillation sources at selected wavelengths. 

Figure \ref{fig:6} shows the inner structure at the height of the magnetic waveguides for the six sunspots in the active groups NOAA 11131 (a), NOAA 11133 (b), NOAA13549 (c), NOAA 13586 (d), NOAA 13314 (e), and NOAA13719 (f).  The lower part shows the white light image of the sunspot at the photospheric level. The upper transparent part is an image of the coronal loops visible at 171~{\AA} wavelength. The volume loops are highlighted where the power of the oscillations is maximized. All waveguides have a similar structure in the form of thin loops of the high-frequency part converging from the sunspot periphery to the center (band III), forming the thickened part of the magnetic bundle (band II) with its stratification, and extending to the penumbra in the form of separate thick waveguides (band I). Their spatial localization coincides with the footpoints of the coronal loops anchored in the umbra and visible at 171~{\AA} wavelength.
    		
		We can see different waveguide distributions, indicating a particular type of magnetic configuration of active groups with varying numbers of open and closed field lines.	The NOAA 11133 group shows a narrow bundle of field lines with a single root in the center of the umbra. This configuration may indicate field line verticality, which is characteristic of small sunspots without a developed trailing part.

		\section{Discussion}
							
					There have been considerable efforts in recent years to explain the observed 3-minute chromospheric umbra oscillations. Two main mechanisms have been proposed. The first is a fast mode resonator, explained by \cite{1981SoPh...71...21S}, formed by increasing Alfven velocity in the photosphere and unstable convection below. This idea has been extended by several researchers, including works of \cite{1993ApJ...402..721C} and \cite{1997SoPh..173..259W}. The second mechanism is the slow mode resonator or sunspot filter approach, localized in the chromosphere and proposed by \cite{1983SoPh...82..369Z}. The resonators are assumed to be located mainly between the transition region and the temperature minimum. Further research followed \cite{1984SvAL...10...19Z, 2018AstL...44..331Z, 2007msfa.conf..351Z,2014AstL...40..576Z,2018RAA....18..105Z}, but it remains unclear which model better describes reality; both may occur in the sunspot atmosphere \citep{1985ApJ...294..682L}.
				
				In our work we have attempted for the first time to obtain experimental evidence for the presence  of oscillation resonators in a sunspot using high spatial and temporal resolution EUV/UV data from SDO/AIA and applying modern methods of spectral signal processing described in \cite{2008SoPh..248..395S} and \cite{2021RSPTA.37900180S}. Three-dimensional images of magnetic waveguides in the volume of the sunspot atmosphere allow us to understand how waves propagate from the sub-photospheric layers into the corona. The spatial structure of the oscillation sources in the form of thin magnetic tubes (Fig.~\ref{fig:3}) anchored in the umbra at the photospheric level shows the formation of magnetic waveguide bundles that diverge and thicken to emerge in the corona. We can see that at all levels of the sunspot atmosphere the variations in EUV/UV emission are a combination of multiple independent oscillations occurring in waveguides with small angular sizes. It can be assumed that each spectral amplitude of the PMD spectra in the detected separate bands of periods I, II, III (Fig.~\ref{fig:2}) corresponds to a different source located at different distances from the umbra center (Fig.~\ref{fig:3}). Their spatial shape varies from point-like and cellular structures in the umbra to elongated and tilted structures in the penumbra. This is in good agreement with the results in \cite{2020ApJ...888...84S}, which show the displacement of oscillation sources at different heights with increasing angular size.
			
			We propose that the observed spatial distribution of wave sources (Fig.~\ref{fig:4}) indicates the set of high-altitude resonators of slow magnetoacoustic waves at the level of the photosphere, chromosphere, and corona, which are part of magnetic waveguides. The amplification of oscillations and the appearance of selected harmonics in the spectrum occur here. The waveguides are formed by the mechanism of the wave cutoff frequency and by the stratification of plasma parameters.  It can be assumed that the inclination of the magnetic field at a certain height in the sunspot atmosphere affects the level of emission. This relationship is nonlinear, depending on the frequency of the wave cutoff, with the peak depending on the cosine of the magnetic field tilt. A similar cosine relationship was previously described in \cite{2014A&A...561A..19Y, 2023SoPh..298...23Y} between the cutoff frequency of magnetoacoustic waves and the magnetic field inclination. 
			
			\subsection{High-frequency oscillations}
						
			High-frequency oscillations with period about 1 minute from sunspot active groups were first detected in the radio band \citep{1984IGAFS..68...21Z, 2004AstL...30..489G}. The emission variations of the radio sources show significant harmonics of quasi-periodic oscillations. The periods are 1.2, 2.8, 4.6 and 5.5 min. It is assumed that oscillations with periods about 1-2 min are harmonics of 5 min periodicity. These type of harmonics was reveal in coronal loops by \cite{2003A&A...404L...1K}. 
			
			In \cite{2018ApJ...856L..16W} shown significant enhancements in the range $\sim$10--14~mHz (about 1 min) from different layers of the sunspot atmosphere.  By comparing the spectra of the sunspot umbra and the distant umbra region, it is shown that the 12~mHz component exists only in the umbra. The propagating wave originates near the footpoints of coronal fan structures anchored near the inner umbra boundary. The velocity of these disturbances near 49 km/s suggests that they are associated with slow magnetoacoustic waves generated by disturbances at the photospheric level of the umbra. \cite{2014A&A...569A..72S} found that the spectra of umbra oscillations contain distinct peaks at 1.9, 2.3, and 2.8 min. Oscillations in the upper atmosphere occur later than in the lower atmosphere. This suggests that the waves propagate from bottom to top.	
			
			\cite{2018A&A...618A.123S} has shown that there are two types of high-frequency periodicity associated with umbral flashes as a background and as a localised phenomenon. The first type, or background UFs, are associated with random yawing of individual parts of interacting wave fronts. These UFs are considered to be weak and diffuse parts that move along the wave fronts without stable shapes and localisation in space. The second type of sources in the form of localised UFs are stable and are mostly observed from the inner part of the umbra boundary.

				We found the presence of high-frequency oscillations on the spectra (band III, Fig.~\ref{fig:2}) which connected with sources of umbral flashes  (Fig.~\ref{fig:6}) in all sunspot groups, without significant period drifts. This indicates a weak vertical variation of the magnetic field in the sources and hence of the cutoff frequency. Upward wave propagation along the magnetic tubes is observed, which in projection to the observer shows the presence  of standing waves.  The main power of these oscillations is concentrated in the lower layers of the sunspot atmosphere and associated with cellular sources located near the inner part of the umbra and forming thin magnetic waveguides \citep{2020ApJ...888...84S}. The presence  of these local sources of high-frequency oscillations at 1700~{\AA} and 1600~{\AA} has been shown in \cite{2019AstL...45..177Z}. The emission variations occur against a background of 3-minute oscillations. The strongest oscillations present at the level of the upper photosphere and the temperature minimum. 
				
				The obtained results about high frequency source localization are consistent with the conclusions on the localization of the sources of the second type of umbral flashes described in \cite{2018A&A...618A.123S}. The volume image of the sources (Fig.~\ref{fig:3}) shows not only the presence  of vertical tubes, but also their expansion and convergence towards the center of the umbra at different heights with the formation of a thick waveguide beam in the II-nd spectral region and strong 3-min oscillations.
		
		 We assume that the detected high-frequency oscillations in the III-rd spectral range are associated with bundles of thin magnetic loops, where the amplification of their separate sections takes place in the form of photospheric resonators. Their occurrence can be explained within the framework of the resonance theory proposed by \cite{1981PAZh....7...44Z} and \cite{1983SoPh...82..369Z}.  It assumes periodic reflections of high-frequency (periods below the 3 min main mode) slow magnetoacoustic waves due to temperature gradients near the photospheric temperature minimum and transition region. These resonators are located downward in a strong vertical field. Wave reflection is not complete; waves can propagate both upward and downward, which determines the scattering of the periods of the trapped waves in the band and the quality of the resonator. We observe such scattering in the form of intensity-period dependence for high-frequency oscillations in PMD spectra (Fig.~\ref{fig:2}) and in the structure of the source shape (Fig.~\ref{fig:3}) as a set of magnetic tubes of different diameters, inclinations, and period bands.   
			
			\subsection{Low-frequency oscillations}
			
					According to \cite{1985SoPh...95...37S}, various changes in physical parameters of the structure of the transition region do not lead to changes in the resonance periods, but only to changes in the amplitudes and phases of the oscillations. Thus, resonant cavities of a certain size and band of oscillation periods are formed, which occupy a part of magnetic waveguides. In \cite{1991ApJ...366..328H} the calculation of the eigenmodes of the sunspot umbra with vertical temperature stratification described by the Maltby model \citep{1986ApJ...306..284M} was performed. The calculations showed the presence  of significant resonant modes in the photosphere with periods of 5 and 3 min, while in the chromosphere only oscillations of slow acoustic modes with a periodicity of 3 min were observed. This is in good agreement with our observations, where at the photospheric level in the 1700~{\AA}~ line, waveguide regions in the form of resonant cavities with 3 and 5 min periods are well distinguished (Fig.~\ref{fig:3}). For chromospheric heights at wavelength 304~{\AA}, the 3-min oscillations become dominant as a thick part of the magnetic bundle of the waveguides. The 5-min oscillations are mostly localized in dedicated waveguides whose size is much smaller than the 3-min part. This greatly reduces the possibility of observing them due to the low power of the oscillations and explains the negative result of the 5-min periodicity registration in \cite{1991ApJ...366..328H}.
	
The obtained images of resonant cavities, covering at different heights the selected bands of oscillation periods, can be described within the framework of the four-layer isothermal model of the sunspot atmosphere developed by \citet{2007msfa.conf..351Z}. In this model, the presence  of four distinct layers with different temperatures corresponding to the corona, the chromospheric temperature plateau, the temperature minimum, and the photospheric-subphotospheric layers has been proposed to study the filtration properties of the atmosphere during wave propagation. We consider a scenario in which there may be an increase in the power of oscillations at the chromospheric level. Slow waves with frequencies below the cutoff frequency of the temperature minimum are strongly reflected and can only penetrate the chromosphere due to the tunneling effect by \citet{1972SoPh...25..329Z}. These waves are then reflected by the corona and penetrate the temperature minimum back into the photosphere. If the waves are in phase, their interference occurs with amplification of the oscillation power, affecting the reflection from the upper atmosphere. The formation of a resonant cavity occurs where slow magnetoacoustic waves are delayed by reflections from the steep temperature gradient in the photosphere and the chromosphere-corona transition zone region, producing the observed spectrum of 3-minute oscillations. This process demonstrates the ability of the sunspot to be as a Fabry-Perot interference filter for slow waves.  Apparently, the spectrum of oscillations in the chromosphere and transition region arises from the combined effects of chromospheric resonance, the cutoff frequency of the temperature minimum, and the nonlinear reflection of the sunspot atmosphere.

It can be assumed that similar interference resonance filters for the amplification of wave oscillations can also operate at other altitude levels.  Their experimental presence  in the form of selected spectral bands related to the sources of oscillations is confirmed by our observations. We suggest that it is necessary to further develop the theory of local helioseismology, with emphasis on the study of the formation of resonant cavities in magnetic waveguides at high levels of the sunspot atmosphere to explain the obtained results.

\section{Conclusions}

For the first time, images of the innner structure of magnetic waveguides and resonant cavities in the sunspot atmosphere have been obtained using local helioseismological methods. The methodology is based on decomposing time cubes of EUV/UV images of sunspots into narrowband components in the form of oscillation sources using pixel wavelet filtering \citep{2008SoPh..248..395S} and obtaining their spectra using the PMD method \citep{2021RSPTA.37900180S}.  

	Based on the study of one-dimensional oscillation period-source coordinate plots for different temperature (height) levels of the observed atmosphere at 1700~{\AA}, 1600~{\AA}. 304~{\AA}, 171~{\AA}, 193~{\AA}, and 211~{\AA} wavelengths, it is shown that different parts of the sunspots show different periodicity of the signals depending on their distance from the sunspot center and the height of the emitting layer.  The one- and two-dimensional structure of the magnetic waveguides in space and height has been constructed and studied in detail for the first time. Separate sections of the waveguides are revealed, where the amplification of the oscillations and the formation of the resonant cavities are observed.	

 Structural changes of the waveguides with height are shown. At the level of the temperature minimum (1700~{\AA}) and the lower chromosphere (1600~{\AA}) three separate bands of periods 1-2 min, 2-3 min and 3-5 min are distinguished, where the maximum power of oscillations is observed. Their appearance is associated with waveguide sections in the form of photospheric and chromospheric resonant cavities. Spectral sources on the period-coordinate plot show an X-shape with the formation of the base in the form of converging thin magnetic waveguides with high oscillation frequency (band III) anchored on the inner boundary of the sunspot umbra. Their footpoints are related to the sources of localised umbral flashes of the second type found in \cite{2018A&A...618A.123S}. As the oscillation period decreases, a broadening of the waveguide size and the formation of a thick magnetic bundle (band II) with a periodicity of 3 min were detected. The penumbra (band I) is characterized by the presence of divergent magnetic structures due to the inclination of the field lines along which running waves propagate. 

It is shown that in the chromosphere at 304~{\AA}, due to the magnetic field output, the shape of the sources changes from X to V-shape and take place formation of a single footpoint anchored in the center of the umbra. The power of the 3-min oscillations is maximal here. Coronal levels are characterized by damping of the oscillations and the appearance of waveguide asymmetry, with waves mostly propagating along open field lines at 171~{\AA}. The source of the 3-min oscillations in the center of the umbra decreases in size and brightness as the waves penetrate along the vertical magnetic field into the hot coronal layers at 193 and 211~{\AA}.

The resonant cavities show their presence at all levels of the sunspot atmosphere. They are most clearly observed in the lower layers at the level of the temperature minimum, which is associated with the maximum power of the oscillations. We can assume that the propagation of waves from subphotospheric levels high into the corona is possible under the condition of their periodic amplification during the passage of resonant cavities at different heights in magnetic waveguides.  The different spectral distributions of the oscillation sources within the sunspot indicate a complex waveguide structure, which is consistent with previous one-dimensional spectroscopic observations.

 The obtained PMD spectra show a similar structure of the oscillation spectra in the form of distinct resonance bands for all investigated sunspot active groups. It is shown that the volume structure of the waveguides can change depending on the magnetic field configuration. This suggests that the field lines along which the waves propagate are the paths where slow magnetosonic waves propagate from the subphotospheric layers into the corona.

 The main reason for the different structure of the waveguides at different heights of sunspots atmosphere is the spatial distribution of the tilt angles of the field lines, and hence the frequency cutoff of the waves. The appearance of volumetric sections of waveguides in the form of resonant cavities is related to the physical parameters of the medium at the places of emission generation.
	
\begin{acknowledgments}
The work was supported by the Ministry of Science and Higher Education of the Russian Federation and Chinese Academy of Sciences President’s International Fellowship Initiative, PIFI Group grant No. 2025PG0008. The authors are grateful to the SDO/AIA/HMI teams for operating the instruments and performing basic data reduction, and especially for their open data policy. 
\end{acknowledgments}

\bibliography{Waveguides} 

\bibliographystyle{aasjournal}

\end{document}